\documentclass[twocolumn,prl,noshowpacs,superscriptaddress]{revtex4}

\usepackage{amssymb}

\usepackage{graphicx}

\begin{document}
\title{Measurement-based quantum computation with superconducting charge qubits}
\author{Xiang-bin Wang}
\affiliation{Frontier Research System, The Institute of Physical
and Chemical Research (RIKEN), Wako-shi 351-0198, Japan}
\affiliation{CREST, Japan Science and Technology Agency (JST),
Kawaguchi, Saitama 332-0012, Japan}
\affiliation{Department of Physics, Tsinghua University, Beijing 100084, China}
\author{J. Q. You}
\affiliation{Frontier Research System, The Institute of Physical
and Chemical Research (RIKEN), Wako-shi 351-0198, Japan}
\affiliation{Department of Physics and Surface Physics Laboratory (National Key
Laboratory), Fudan University, Shanghai 200433, China}
\author{Franco Nori}
\affiliation{Frontier Research System, The Institute of Physical
and Chemical Research (RIKEN), Wako-shi 351-0198, Japan}
\affiliation{Center for Theoretical Physics, Physics Department,
Center for the Study of Complex Systems,
University of Michigan, Ann Arbor, MI 48109-1040, USA}

\begin{abstract}
We present a robust method, based only on measurements, to produce 
superconducting cluster states. The measurement of the current of a few
parallel Josephson-junction qubits realizes a novel type
of quantum-state selector. Using this selector, one can produce
various quantum entangled states and also realize a controlled-NOT
gate without requiring an exact control of the interqubit interactions. 
In particular, cluster states for quantum computation could be produced
with only single-qubit measurements.
\end{abstract}
\maketitle {\it Introduction.---\/}
%
Joint operations of two qubits are crucial for quantum information
processing. Indeed, in principle, controlled-NOT (CNOT) gates and single-qubit
unitary transforms are sufficient for quantum computing. However,
implementing a CNOT gate via two interacting qubits has proven to be extremely
difficult. This is a huge barrier to scalable quantum
computing, which requires numerous CNOT gates.

 To avoid the daunting difficulties related to controllable-interaction-based 
CNOT gates, several attempts have been made towards the goal of doing quantum
computation {\it without} these and based on entangled states and
measurements only \cite{pittman0,knill,gottc,others}. Indeed, it
is possible to replace CNOT gates by quantum teleportation
\cite{gottc}, where the only collective operation is a Bell
measurement.
However, a complete Bell
measurement is also a very challenging task. Moreover, 
a {\em probabilistic} CNOT gate through Bell measurement, demonstrated
experimentally with an optical set-up \cite{pittman0}, 
cannot be directly used for
practical large-scale quantum computation \cite{pittman0}.
A very elegant alternative is given by one-way quantum computation using
highly entangled states called cluster states~\cite{cluster1}. This method  
has been proposed as a potential way to 
solve the very challenging problems faced by standard approaches to 
quantum computing. Cluster
states must be first produced and stored as a ``resource" to be
consumed later for quantum computation through individual
measurements only. The first important step here
is to generate the so-called cluster states.
Naively, one can achieve this goal through CNOT gates or
conditional phase shift (CPhase) operations, and these operations
could be obtained if we had good control of 2-qubit interactions.
However, done this way, the advantages of cluster-state quantum
computation become weak.

Cluster-state quantum computation has recently been demonstrated
in a quantum optical experiment \cite{zei} through non-deterministic
Bell measurements. However, an optical quantum computing set-up
also has its own inevitable disadvantages. For instance, it is
difficult to store an optical quantum state for future use. Also,
due to the post-selection nature of the result, it is quite
difficult to perform scalable optical quantum computing using cluster
states.
To overcome these drawbacks, one could either try to improve the
technology of optical systems or consider another system that
could produce cluster states: (1) {\it without} post-selection, and (2) that
could be {\it stored} for future use. Here we consider this later approach
using Josephson-junction (JJ) circuits
\cite{today,liu,wei,vion,you0,you,pash}. We shall present a
method to produce cluster states with superconducting qubits
through the mechanism of quantum state selection.
Remarkably, this method does not require any precise control of either
the interqubit interactions or the timing. Since this approach can
produce cluster states without post-selection, the cluster states
can be stored as a resource for performing measurement-based
quantum computing \cite{cluster1}.

{\it Quantum-state selection with charge qubits.---\/} We consider a
circuit with one large junction denoted by ``0" and many parallel
charge qubits made up of smaller junctions, as shown in Fig. 1. 
If the current across junction 0 is larger than a certain critical value
$I_{T0}$, it switches from the superconducting state to the normal
(resistive) state. Usually, the current contributed from those smaller
junctions is significantly less than $I_{T0}$. With an appropriate
bias current, the current from qubits 1 to $k$ determines whether
the large junction 0 will be switched to the resistive
state with a nonzero voltage $V$. The current is determined by the
quantum state of those small JJ qubits in the circuit. Therefore,
by monitoring the voltage $V$, one can determine which type of state
those JJ qubits have been projected to~\cite{vion}.

\begin{figure}
\includegraphics[width=3.4in,
bbllx=10,bblly=408,bburx=582,bbury=689]{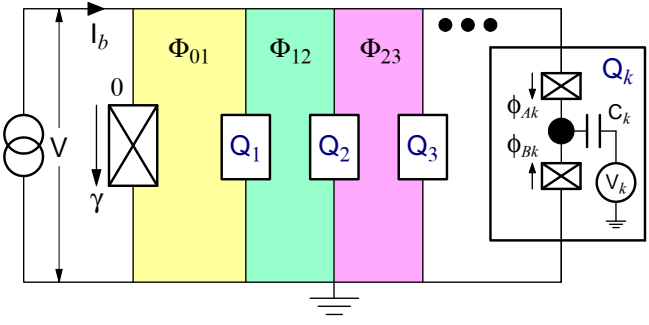} 
\caption{\label{Fig1} (Color online) 
A Josephson-Junction circuit with one large
junction ``0" and many parallel charge qubits ($Q_1\sim Q_k$). Each qubit
consists of two small junctions. The detailed structure of each
qubit is schematically shown in $Q_k$. Qubit $Q_i$ is decoupled when the applied 
flux $\Phi_{0i}\equiv \sum_{j=0}^{i-1} \Phi_{j,j+1}$  is tuned to be zero. 
Based on this circuit, one can
perform quantum state selection on any subset of qubits. Also, one can
generate an entangled pair state on any two qubits and produce a CNOT gate
on any two qubits with one ancilla. Moreover, one can produce a cluster
state for one-way quantum computation.}
\end{figure}

We shall use the following notation. ${\cal L}_{ij}$: the loop
connecting junctions $i$ and $j$, $s_{ij}$: the region enclosed by
loop ${\cal L}_{ij}$, $\Phi_{ij}$: magnetic flux threading the
region $s_{ij}$. In Fig.~1, $I_b$ is the bias current, $I_b$
and $\Phi_{ij}$ can be tuned, and $\Phi_0$ is the flux quantum.
There are two sets, $S$ and $\bar S$, of states for those observed
qubits. A state in set $S$ ($\bar S$) will (will not) cause 
junction 0 to switch from the superconducting to the normal
state, given a certain bias current $I_b$ and external fields.
Thus, we can conclude whether the
quantum state of those observed qubits belongs to the set $S$ or 
$\bar S$, by just monitoring the voltage $V$. This process can be
regarded as a type of incomplete {\it measurement}, a measurement which
only projects the observed system to a subspace rather than a 
single state.  If $V=0$, junction ``0" must still be in the
superconducting state; therefore, the projected quantum state
$\Psi_{1\cdots k}$ (which belongs to $\bar S$) 
for qubits $\{1,2,\cdots, k\}$ must satisfy
\begin{equation}\label{ine1}
\left| I_b+ \langle \Psi_{1\cdots k}|\hat I |\Psi_{1\cdots
k}\rangle\right| < I_{T0}.
\end{equation}
Here $\hat I$ is the current operator for those qubits. If
$V\not=0$ is detected, the projected state (now belonging to $S$) must satisfy
\begin{equation}\label{ine2}
\left| I_b+ \langle \Psi_{1\cdots k}|\hat I |\Psi_{1\cdots
k}\rangle\right|
> I_{T0}.
\end{equation}
These two conditions realize a {\it quantum-state selector}. Let us
now assume an initial state $\Psi_{1\cdots k}=\alpha |a\rangle
+\beta |b\rangle$. Suppose $|a\rangle\in \bar S$ and $|b\rangle\in
S$, i.e., state $|a\rangle$ and $|b\rangle$ satisfy the conditions
$I_b+\langle a|\hat I|a\rangle < I_{T0}$ and $I_b+\langle
b|\hat I|b\rangle > I_{T0}$. With the bias current $I_b$, the
quantum state for $k$ qubits must be $|a\rangle$ if $V=0$. This
fact helps to prepare various types of entangled states including
cluster states, as we show below.

If we set $I_b$ to be significantly smaller than the critical
current, the total current across junction 0 will be very small
and thus $V=0$ for whatever state of the small JJ qubits in the
circuit. Thus, if the bias current $I_b$ is very small, there is no
measurement. But if a bias current $I_b$ slightly below $I_{T0}$ is
applied, the state of those coupled qubits determines whether
the large junction 0 switches to a normal state. Therefore, the state of
those coupled qubits can be  measured via $V$, after applying an
appropriate $I_b$.

Actually, we can also make the quantum state selection to any
subset of those qubits in the circuit. As we shall show, by
applying appropriate flux in each region, we can select a
subset of qubits which provide a negligible contribution to the total
current. This means we can decouple some qubits by tuning the
external field.

Consider loop ${\cal L}_{0j}$ with flux $\Phi_{0j}$, which contains the 
large junctions 0 and the charge qubit $j$.
The following constraint holds for the
phase operators across the junctions 0 and those in qubit $j$: 
$\hat \phi_{Aj}-\hat \phi_{Bj}-\hat \gamma  +
{2\pi\Phi_{0j}}/{\Phi_0}=0 $ and $\Phi_{0j}=\sum_{i=0}^{j-1}
\Phi_{i,i+1}$, where $\Phi_{i,i+1}$ is the flux threading the region
$s_{i,i+1}$ and $\hat \gamma$, $\hat \phi$\,'s the JJ phase drops (see Fig.~1). 
The total current across junction 0 is given by~\cite{you}
\begin{equation}
I_{c0}\sin \hat \gamma  = 2 \sum_{j=1}^k \sin
\left(\frac{\pi\Phi_{ej}}{\Phi_0}-\frac{1}{2}\gamma\right)I_{cj}\cos\hat
\phi_j
\end{equation}
and $\phi_j=(\phi_{Aj}+\phi_{Bj})/2$, $I_{cj}$ is the critical
current of the junction in qubit $j$. Since junction 0 is large,
$\langle\hat\gamma\rangle$ must be small. Therefore we can
generalize Eq.~(19) in Ref.~\cite{you} (from two to $k$ qubits) for the total 
current operator $\hat I$ of all parallel qubits:
\begin{eqnarray}
\hat I \!&\!=\!&\! \sum_{j=1}^k
\left[\sin\left(\frac{\pi\Phi_{0j}}{\Phi_0}\right)
\right] I_{cj}\sigma_{x}^{j}-C \nonumber \\
&&-\frac{1}{{2I_{c0}}} \sum_{j> i\ge 1}
\sin\left[\frac{\pi(\Phi_{0i}+\Phi_{0j})}{\Phi_0}\right]
I_{ci}I_{cj}\sigma_{x}^{i}\sigma_x^j.
\end{eqnarray}
Here  $C=({1}/{4I_{c0}})\sum_{j=1}^k \sin
({2\pi\Phi_{0j}}/{\Phi_0})I_{cj}^2$. 
To decouple any qubit $i$, external fields must be tuned so that
$\Phi_{0i}=0$.

{\it Producing two-qubit entanglement.---\/} We first consider the
simplest application for generating two-qubit entangled
states, e.g, $|\psi^{\pm}\rangle=\frac{1}{\sqrt
2}(|01\rangle \mp |10\rangle)$. Here,
$|0\rangle$, $|1\rangle$ represent the state of 0, 1 extra Cooper
pair, respectively.
Consider now opposite external magnetic fields in regions $s_{01} $
and $s_{23}$, i.e., $\Phi_{01}=-\Phi_{23}=\Phi_0/2$. Therefore,
given any $i$, $\Phi_{0i}=0$ if $i>2$, $\Phi_{0i}=\Phi_{0}/2$ if
$i=1$ or $i=2$. Only qubits 1 and  2 would contribute to the total
current. Initially, the state of qubits 1 and 2 is set to be
\begin{equation}
|\Psi_{1,2}\rangle =|0\rangle|0\rangle=\frac{1}{2}(|+\rangle +
|-\rangle)(|+\rangle + |-\rangle).
\end{equation}
Here, the first (second) state is for qubit 1 (2).
 Note that states
 $ |\pm\rangle = \frac{1}{\sqrt 2}(|0\rangle \pm |1\rangle)$
 are eigenstates of $\sigma_x$ with $\sigma_x |\pm\rangle =\pm|\pm\rangle$.
We then bias the large junction and monitor $V$.
Here we choose
$I_b= I_{T0}-(I_{c1}+I_{c2})/2$.
 If $V=0$, the state $|+\rangle|+\rangle$ is ruled out and the
state must have been collapsed to
$|+\rangle|-\rangle + |-\rangle|+\rangle +|-\rangle|-\rangle.
$
We then reverse both, the external fields and bias current, to their opposite 
directions and again monitor $V$. If $V=0$ again, 
we conclude that we have now prepared the entangled state $|\psi^-\rangle$.
The success probability is 1/3. If at any stage we find $V\not=0$,
the protocol fails and we need to redo it from the beginning.
If we had initially used $|0\rangle|1\rangle$, after the above
quantum state selection, we would be obtaining $|\psi^+\rangle$. A
similar idea has been proposed to prepare a two-qubit entangled
state with weak continuous measurement \cite{rr}. Another proposal
to generate entangled states with controllable interactions was
raised very recently \cite{wei}.

{\it Four-qubit cluster state.---\/} To produce a nontrivial
cluster state (with four qubits), a  CNOT gate must be applied on
two entangled pairs.
Remarkably, such type of CNOT gate can also be done with our
quantum state selector. Here we follow the approach proposed by
Pittman et al \cite{pittman0}. Consider three qubits. Qubits 1 and 3
are the control and target qubits, respectively, and qubit 2 is the
ancilla. Initially, qubit 2 is in state $|+\rangle$. Without any loss
of generality, we can consider the following initial state for the
three qubits:
\begin{equation}
|\psi_0\rangle=|i\rangle|+\rangle|j\rangle
\end{equation}
where the first state is for the control qubit, the second and
third are for the ancilla and target qubits, respectively, and
$i$, $j$ can be either 0 or 1. A CNOT gate is obtained by the
following operations: (1) Measure the parity value of the ancilla  and
control qubits in the $|0,1\rangle$ basis (parity=1  for the subspace
$\{|01\rangle,|10\rangle\}$);
(2)measure the parity value of the ancilla and target qubits in
the $|\pm\rangle$ basis (parity =1  for space
$\{|+-\rangle,|-+\rangle\}$); (3) perform a Hadamard transform to the
ancilla; (4) measure the ancilla in the $|\pm\rangle$ basis; (5) perform
an individual unitary transformation on the target qubit according to the
previous measurement result of the ancilla. In particular, if we
obtain 1 in both parity measurements in the first two steps, the
following state was prepared (up to a phase factor which can be
removed by an individual phase-shift operation):
\begin{equation}
|\psi_1\rangle = \frac{1}{\sqrt 2}( |i\rangle |+\rangle|i\oplus
j\rangle + |i\rangle |-\rangle | 1\oplus i \oplus j\rangle ).
\end{equation}
Thus, if the state $|+\rangle$ is detected for the ancilla, a CNOT
gate was applied to the control and target qubits;  if state
$|-\rangle$ is detected for the ancilla, a CNOT gate is applied
after flipping the target qubit in the $|0,1\rangle$ basis.

The scheme above can be realized by the JJ quantum state selector.
Consider figure 1 for a circuit of charge qubits.  Here qubit 2 is
regarded as the ancilla, while qubit 1 and 3 are the control and the
target, respectively. To measure the parity of qubits 1 and 2,
qubit 3 must be decoupled first. This can be done by applying
$\Phi_{01}=-\Phi_{23}=\Phi_0/2$ and $\Phi_{12}=0$.  The quantum
state selection operation described above is in the $|\pm\rangle$ basis.
To measure the parity value in the $|0,1\rangle$ basis, the following
operations are needed: (a) Prepare the initial state of the
ancilla in $|0\rangle$; (b) perform a Hadamard transform to qubit 1; 
(c) activate the quantum state selection operation so that there must
be one $|+\rangle$ state and one $|-\rangle$ state for qubits 1
and 2;  (d) perform a Hadamard transform to qubits 1 and 2,
respectively. Later, the parity of qubits 2 and 3 must be measured in
the $|\pm\rangle$ basis.  An external field in region $s_{12}$ only
needs to be applied to decouple qubit 1.  The following operations
suffice for the measurement: (a) Apply an appropriate bias current
and do quantum state selection to qubits 2 and 3 so that there are
one $|+\rangle$ and one $|-\rangle$ states;  (b) apply a Hadamard
transform on qubit 2;  
(c) measure qubit 2 in the $|\pm\rangle$ basis. Given the outcome
$|+\rangle$ for qubit 2, a CNOT gate was generated for qubit 1 and
3; given the outcome $|-\rangle$, a CNOT gate was generated after
flipping qubit 3 in the $|0,1\rangle$ basis.

Note that although we need to tune the external fields for our
CNOT gate, we do not need to control them and their timing {\em
exactly}. This is an essential difference from a CNOT gate
through controlling the qubit-interacting time. A probabilistic CNOT
gate~\cite{pittman0} itself cannot be directly applied for scalable quantum
computing. However, it is useful in our picture of
cluster-state quantum computation with quantum state selection: At
a convenient time, using the superconducting quantum state
selector, we can generate cluster states and
store them as a resource. In making a computation, the cluster states,
prepared in advance, are consumed.

The circuit in Fig.~1 (with five qubits ) can generate a four-qubit cluster state 
(qubits 4 and 5 are not explicitly drawn in Fig. 1). 
Our goal is to prepare a cluster
state on qubits 1, 2, 4 and 5. Qubit 3 is the ancilla for the CNOT
gate on qubits 2 and 4. We can first prepare two pair states
$|\psi^-\rangle$ on qubits 1 and 2 and on qubits 4 and 5, respectively, 
and then apply a probabilistic CNOT gate through quantum state
selection on qubits 2 and 4, to produce a cluster state. The
detailed procedure is now given: (1) Set $\Phi_{01}=-\Phi_{23}=\Phi_0/2 $ and
$\Phi_{12}=\Phi_{34}=\Phi_{45}=0$. Qubits 3, 4, and 5 are now
decoupled.  Apply an appropriate bias current and observe $V$.
Then, use opposite fields and bias current and observe $V$
again. If $V$ is always 0,  the quantum state selection process
has functioned successfully, and the state of qubits 1 and 2 has been
projected to
$|\psi^-\rangle =\frac{1}{\sqrt 2}(|01\rangle-|10\rangle)$.   
(2) Similarly,  the pair
$|\psi^-\rangle$ state can also be prepared on qubits 4 and 5. To
do so, the external field should only be applied in region $s_{34}$, 
so that only qubits 4 and 5 can contribute to the total
current, while qubits 1, 2 and 3 are all decoupled.  
(3) Make a CNOT gate on qubits 2 and 4. A cluster state of the form
\begin{equation}\label{cc}
|\psi_c\rangle= \frac{1}{\sqrt 2} \left(|01\rangle|\phi^-\rangle
-|10\rangle |\psi^-\rangle\right)
\end{equation}
is prepared for qubits 1, 2, 4 and 5, and
$|\phi^-\rangle=\frac{1}{\sqrt 2}(|00\rangle-|11\rangle)$. Note
that by individual unitary transforms this state is equivalent to
$\frac{1}{2}(|0000\rangle+|0011\rangle+|1100\rangle-|1111\rangle)$,  
which is used in Ref.~\cite{zei}. To make the CNOT gate, the parities of
qubits 2 and 3 in the $|0,1\rangle$ basis and qubits 3 and 4 in
the $|\pm\rangle$ basis must be measured, as stated earlier. To
measure the parity of qubits 2 and 3, qubits 1, 4 and 5 should be
decoupled by setting $\Phi_{12}=-\Phi_{34}=\Phi_0/2$ and
$\Phi_{01}=\Phi_{23}=\Phi_{45}=0$. 
Moreover, by setting $\Phi_{23}=-\Phi_{45}$ and
$\Phi_{01}=\Phi_{12}=\Phi_{34}=0$, qubits 1, 2 and 5 are decoupled
and the parity of qubits 3 and 4 can be measured through the
quantum state selector. If $V=0$ in both parity measurements, we
have prepared the following state for the five qubits:
$$
|\Psi_{1\cdots 5}\rangle=\frac{1}{\sqrt 2}
\left[|+\rangle_3\left(|01\rangle_{1,2}|\phi^-\rangle_{4,5}
-|10\rangle_{1,2}|\psi^-\rangle_{4,5}\right)\right.
$$
\begin{equation}
+\left.|-\rangle_3\left(|01\rangle_{1,2}|\psi^-\rangle_{4,5}
-|10\rangle_{1,2}|\phi^-\rangle_{4,5}\right)\right]
\end{equation}
A cluster state on qubits
1, 2, 4 and 5 is readily obtained after measuring  qubit 3 in the
$|\pm\rangle$ basis.

{\em Efficiency.---\/} We have shown above that cluster states can be
produced probabilistically using only measurements. Our
pre-selection result here is totally different from the post
selection result or  probabilistic quantum computing in an optical
set-up. In an optical set-up with post-selection, the state is
destroyed after a measurement.  However, in our approach here,
one can first verify the state via a measurement (the state is
still there after the measurement) performed at a convenient time and then
do the computing when needed. 

Suppose we want a large cluster
state containing $N$ qubits. Such a large cluster state is
sufficient to treat a complex task, e.g., factoring a huge
number larger than $2^{\alpha N}$, where $\alpha\sim 0.1$. 
In our approach, we can make this in a concatenated
manner: We first produce many small identical qubits blocks
(pairs), then generate a larger block by combining two small
qubits blocks via a CPhase or CNOT gate, where the target is from
one small block and the control is from another small block. After
each successful combination, the size of the state is doubled. If
the CPhase or CNOT gates are continuously successful $n$ times, we
have constructed a cluster state of size $2^n$. To generate a
cluster state containing more than $N$ qubits, we need $n=\log_2
N$. Therefore, in our approach the joint probability to produce a
cluster state containing more than $N$ qubits is a linear
function of $N^{-1}$. This is only {\em polynomially} small. Now
consider generating such a state using an optical set-up with
post-selection: One first produces $\frac{1}{2}N$ photon pairs and then
generates entanglement using $N$ beam-splitters. If all beam-splitters
have functioned correctly (e.g., one photon on each side of the
outside ports of a beam-splitter), the expected cluster state is
produced. In such a way, the final success probability is a tiny
$2^{-N}$ ({\em exponentially} small).
Note that in an optical set-up with
post-selection, one cannot iteratively make a cluster state block
by block, because there is no way to verify the state of a certain
individual block without destroying it. To overcome this drawback,
one can either develop optical technologies
so that a photon can be measured without damage and can be stored,
or use our approach with solid state qubits.

{\em Concluding remarks.---\/}
We have shown how to make the cluster states through a mechanism
of quantum-state selection with charge qubits.  Obviously, besides the
four-qubit cluster state, our method can also be applied to
generate {\em Multi}-qubit cluster states since our method is based
on a scalable circuit. Moreover, our method also applies for other
solid-state systems, e.g., quantum dot charged qubits~\cite{TT}. 
It is important to stress that, in our scheme, we do not have to control
the interqubit interaction and their timing, e.g., there is no need to worry
about small errors in the control of the external fields,
provided that they do not violate the inequalities
(\ref{ine1}) and (\ref{ine2}). Our result is also robust with respect to
small errors in the control of the bias current $I_b$. Besides cluster
states, the powerful quantum-state selector method presented here can also 
be used to produce many other types of entangled states, including GHZ states
and $W$ states. 


{\bf Acknowledgement:} 
This work was supported in part by the NSA, LPS, ARO, NSF and NSFC.

\end{document}